# Evaluating performance of hybrid quantum optimization algorithms for MAXCUT Clustering using IBM runtime environment

Daniel Beaulieu[1] and Anh Pham[2]

[1]Deloitte Consulting LLP, Arlington, VA 22209
[2]Deloitte Consulting LLP, Atlanta, GA 30303

**Quantum algorithms can be used to perform unsupervised machine learning tasks like data clustering by mapping the distance between data points to a graph optimization problem (i.e. MAXCUT) and finding optimal solution through energy minimization using hybrid quantum classical methods. Taking advantage of the IBM runtime environment, we benchmark the performance of the "Warm-Start" (ws) variant of Quantum Approximate Optimization Algorithm (QAOA) versus the standard implementation of QAOA and the variational quantum eigensolver (VQE) for unstructured clustering problems using real world dataset with respect to accuracy and execution time. Our numerical results show a strong speedup in execution time for different optimization algorithms using the IBM Qiskit Runtime architecture and increased speedup in classification accuracy in ws-QAOA algorithm.**

## 1 Introduction

Unsupervised learning is an important methodology used by data scientists to find emergent properties of data without requiring labelled data or specifying which variable in the data set to make predictions based on. Clustering is a important unsupervised methodology for understanding scientific and consumer data. Clustering has a wide range of applications including exploratory data analysis, reducing dimensionality, and image processing. Quantum optimization algorithms can be used to perform unsupervised clustering tasks, and cluster data in a manner similar to regularly used classical clustering algorithms like K-means on noisy intermediate-scale quantum (NISQ) devices [1], [2]. Specifically, by transforming the distance between data points to weighted graph problem like MAXCUT, binary classification can be performed by finding the groundstate using the quantum approximate optimization algorithm (QAOA)[2]. However, the performance of QAOA is strongly dependent on graph topology [3] and edge weights [4] making QAOA unsuitable to solve complex optimization problems[3],[4]. Harrigan et al. found that QAOA performance declines for circuits with increasing circuit depth due to noise on NISQ hardware[3].

An alternative algorithm to conventional QAOA is the warm-start QAOA (ws-QAOA)[5]. Within ws-QAOA, the initial state and the mixer Hamiltonian can be derived from the solutions of the relaxed binary constrained quadratic (BQM) equations. As a result, ws-QAOA is expected to produce better results than conventional QAOA. In addition, other hybrid algorithm like the variational quantum eigensolver (VQE) can also be applied to solve the MAXCUT problem.

In this paper, we benchmark the performance of different quantum optimization algorithms to solve the MAXCUT problem using weighted graphs derived from a real dataset. We tested the normal QAOA, VQE, ws-QAOA using the Qiskit Runtime API implementations of these hybrid algorithms on a classical simulator and real quantum devices (IBM Lagos and IBM Nairobi) using 8192 shots. Our numerical results reveal the applicability and limitation of these optimization algorithms in performing clustering on real world datasets.

Daniel Beaulieu: dabeaulieu@deloitte.com
Anh Pham: anhdpham@deloitte.com

## 2 Methodology

### 2.1 Engineering dataset into MAXCUT problem

To apply hybrid quantum algorithms for data clustering, we first calculate the Euclidean distance between different data points as following

$$w_{ij} = ||x_i - x_j|| \quad (1)$$

where $w_{ij}$ is the weight matrix representing the distance between the two different data points in a set of N points. This weight matrix is now mapped to a MAXCUT problem where the values of $w_{ij}$ represent the edge length connect the two nodes. Within this formalism the cost function is defined as an Ising Hamiltonian

$$H_C = -\sum_{i,j=1}^{\infty} \frac{w_{ij}}{2}(1 - \sigma_i^z \sigma_j^z) \quad (2)$$

By minimizing the energy of this Hamiltonian we can obtain the optimal solution in classical binary bits which represent the result of our binary classification. In this study, different hybrid quantum optimization algorithms (VQE [6] [7], QAOA [8], and ws-QAOA [5] [9]) were used to find the optimal bit string for the weighted MaxCut instance. The classical optimizer used in all our algorithm was SPSA due to its robustness against noise in NISQ hardware [10]. All the results were also compared against the solutions obtained from the qiskit exact diagonalization.

### 2.2 Normal QAOA

In the normal QAOA algorithm, a specific ansatz is used to approximate an annealing process. This ansatz is defined as a product between the unitary operators constructed from a mixer Hamiltonian $H_B = \sum_j^N \sigma_j^x$ and the cost Hamiltonian $H_C$. The optimal solution is obtained by finding the parameters in the ansatz which minimizes the expectation value $\langle \Psi_{\beta,\gamma} | H_C | \Psi_{\beta,\gamma} \rangle$

$$\Psi(\beta, \gamma) = \prod_{\alpha=1}^{p} e^{-i\beta_\alpha H_B} e^{-i\gamma_\alpha H_C} |+\rangle^N \quad (3)$$

In equation (3), p is the repetition time, and $|+\rangle^n$ is the ground state of the mixing Hamiltonian $H_B$.

### 2.3 Warm-Start QAOA

The conventional QAOA has lacked theoretical guarantees on its performance as well as its ability to outperform classical algorithms [11] [4]. The idea of warm-start QAOA is to use a continuous-valued relaxation that is positive and semi-definite to relax a convex quadratic program and find an optimal initial starting point for QAOA [5] [9]. Warm-start QAOA has been shown to have a higher probability of sampling the optimal solution [5].

Within the formalism of ws-QAOA, the mixer Hamiltonian and the ground state can be constructed as

$$H_{B_i}^{ws} = \begin{pmatrix} 2c_i^* - 1 & -2\sqrt{c_i^*(2c_i^* - 1)} \\ -2\sqrt{c_i^*(2c_i^* - 1)} & 1 - 2c_i \end{pmatrix} \quad (4)$$

$$|GS\rangle = \otimes_{i=0}^{n-1} R_y(\theta_i)|0\rangle_n \quad (5)$$

$$\theta_i = 2arcsin(\sqrt{c_i^*}) \quad (6)$$

where $c_i^*$ is the solution of the relaxed QUBO formulation as described below

### 2.4 Warm-Start QAOA Process

- Transform the problem data into a QUBO quadratic formulation

- Relax the QUBO formulation and its binary variables into a semi-definite convex problem which can be solved using a classical optimizer so it becomes a multiple continuous variables with a range of 0 to 1

- Convert problem to an Ising Hamiltonian

- Pass the relaxed semi-definite continuous variables to form the mixer operator for the Warm-started QAOA

- Create a mixer operator

Due to the numerical results of the weight matrix our QUBO becomes non-convex. As a result, we could not use the CPLEX optimizer to solve our relaxed semi-definite QUBO quadratic program, but instead we ultilized a more general solver, e.g. the GuRoBi solver, which does not require convex quadratic programs. The solution from the relaxed QUBO problem was then used to effectively warm start QAOA. While CPLEX is found to perform better under high-dimensionality problems [12], GuRoBi was able to produce results with the non-convex quadratic problem created from our data.

## 2.5 Varitation Quantum Eigensolver (VQE)

For our VQE algorithm to perform the optimization of the Ising Hamiltonian, we implemented a variational quantum circuit compose of the $R_y$ gates and the two qubit controled CNOT gates. This minimal circuit was then repeated 5 times to create a parameterized circuit. This Ansatz is created by the Qiskit TwoLocal parameterized circuit from the Qiskit Circuit Library[13]. The single qubit rotational angles were optimized using the SPSA classical optimizer to obtain the minimum energy of the expectation value of the cost Hamiltonian. This Ansatz is expected to represent the lowest energy solution possible for VQE.

In addition, we also tested the repetition of the ansatz from length of 1 to length of 5 to investigate the dependence optimization result on the circuit depth. The results on the simulator indicated that using the TwoLocal ansatz cannot produce the optimal solution regardless of the repetition length of the circuit. As a result, here we only presented the solution with a circuit length of 5 in the result section

## 2.6 Data

We performed the binary classification on two different data sets. In the first data set, we used the 1974 Motor Trend US magazine[14]. The clustering objective was to cluster cars that were either sports cars or economy cars. Only five rows were selected from this data set to reduce the number of jobs that would need to be submitted to quantum back-end hardware. We also tested an additional wine data set with 6 rows of dataset to understand the validate our methodology with higher number of qubits. The wine data came from the UCI Machine Learning Library and are based on red and white vinho verde wine samples from north Portugal[15].

## 3 Results

Our optimization results were done using 8192 shots for the real quantum hardware and qasm simulator

## 3.1 Simulator Results

### 3.1.1 Base Quantum Optimization Algorithm Results

The results show that QAOA and VQE run in the normal manner did not produce accurate clustering results compared to the ground truth in the real data or the output of the classical optimization algorithm. Normal QAOA in particular had a much higher energy than VQE or the compared to VQE or the classical algorithm implying worse performance as shown in convergence plot in Figure 2 .

When run on the seven qubit IBM Lagos system, neither QAOA or VQE assigned all observations to the correct cluster. However, also Base VQE and QAOA did not correctly assign observations to the correct cluster on the simulator, so the task performance wasn't worse on quantum hardware. When run on quantum hardware QAOA and VQE both incorrectly clustered observations. The solution objective was lower for both QAOA and VQE on quantum hardware than on the simulator, which can be read as a rating of lower performance on quantum hardware. The process time for base VQE and QAOA was significantly longer than on the simulator, which is an expected result.

### 3.1.2 Qiskit Runtime Quantum Optimization Algorithm Results

For the Qiskit Runtime versions of the algorithm, VQE produced incorrect clustering results, and QAOA produced correct results, albeit with a low energy. The Qiskit Runtime versions were not expected to yield time-savings on simulator hardware, and did not run the processes faster than the non-Qiskit Runtime optimization versions of the algorithms.

The Qiskit Runtime versions of QAOA returned the same, and correct answer as the classical method. VQE returned errors when run with Qiskit Runtime on actual quantum hardware, although the non-runtime version of VQE also did not assign all records correctly. The solution objective for both VQE and QAOA when run on Qiskit Runtime as zero, which is an unexpected result.

The advantage that warm-start QAOA has over normal QAOA is demonstrated clearly in Figure 3. The normal QAOA shows multi-ple peaks in the Eigenstate vector energy graph which could be chosen as an optimal. The warm-start QAOA only has one clear peak in Eigenstate vector energy, showing a clearer result. The results for warm-start QAOA are more comparable to VQE, which also shows a pronounced Eigenstate vector energy peak.

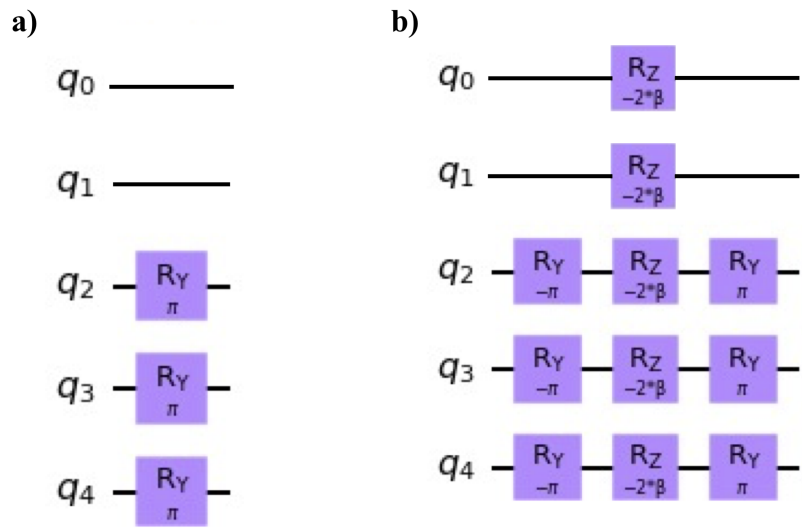

Figure 1: Quantum circuit for ws-QAOA obtained from the car dataset. a) Initial state. b) ws-mixer Hamiltonian

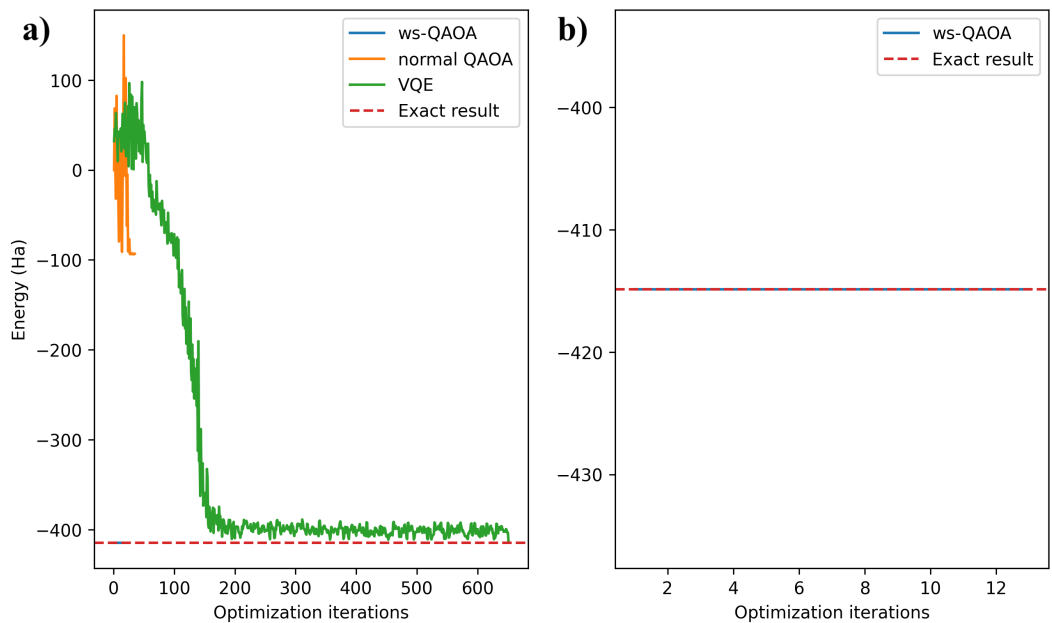

Figure 2: Convergence steps for different quantum optimization algorithms on simulator. a) Convergence for ws-QAOA, normal QAOA, VQE and exact result. b) Zoom-in convergence steps for ws-QAOA

Table 1: Unsupervised Clustering Results by Different Algorithm on Simulator

| Simulator | Type | Classical | VQE (Normal) | QAOA (Normal) | ws-QAOA (Normal) |
|---|---|---|---|---|---|
| Honda Civic | economy | 1 | 0 | 1 | 0 |
| Toyota Corolla | economy | 1 | 0 | 1 | 0 |
| Camaro Z28 | sport | 0 | 1 | 0 | 1 |
| Pontiac Firebird | sport | 0 | 0 | 1 | 1 |
| Maserati Bora | sport | 0 | 1 | 0 | 1 |
| Energy (Ha) | | -414.86 | -410.912 | -117.996 | -414.857 |
| Solution Objective | | 1163.497 | 1159.314 | 1159.314 | 1163.497 |
| Process time (s) | | 0.014 | 4.195 | 0.263 | 0.356 |

Table 2: Unsupervised Clustering Results by Different Algorithm on Simulator with Qiskit Runtime Environment

| Simulator | Type | Classical | VQE (Runtime) | QAOA (Runtime) | ws-QAOA (Runtime) |
|---|---|---|---|---|---|
| Honda Civic | economy | 1 | 0 | 1 | 0 |
| Toyota Corolla | economy | 1 | 0 | 1 | 0 |
| Camaro Z28 | sport | 0 | 1 | 0 | 1 |
| Pontiac Firebird | sport | 0 | 1 | 0 | 1 |
| Maserati Bora | sport | 0 | 0 | 0 | 1 |
| Energy (Ha) | | -414.86 | -411.455 | -11.640 | -414.857 |
| Solution Objective | | 1163.497 | 859.489 | 0 | 1163.497 |
| Process time (s) | | 0.014 | 1.606 | 1.508 | 1.0119 |

Table 3: Unsupervised Clustering Results by Algorithm on IBM Lagos

| Simulator | Type | Classical | VQE (Normal) | QAOA (Normal) | ws-QAOA (Normal) |
|---|---|---|---|---|---|
| Honda Civic | economy | 1 | 0 | 1 | 0 |
| Toyota Corolla | economy | 1 | 0 | 1 | 0 |
| Camaro Z28 | sport | 0 | 1 | 0 | 1 |
| Pontiac Firebird | sport | 0 | 0 | 1 | 1 |
| Maserati Bora | sport | 0 | 1 | 0 | 1 |
| Energy (Ha) | | -414.86 | -114.168 | 13.381 | -164.519 |
| Solution Objective | | 1163.497 | 1159.313 | 859.49 | 1163.497 |
| Process time (s) | | 0.014 | 174.453 | 25.031 | 81.178 |

### 3.1.3 Warm-start QAOA Algorithm Results

Warm-start QAOA produced accurate clustering results with low energy levels as good as the classical algorithm as shown in Figure 2. The warm-start QAOA algorithm produced the best unsupervised clustering results of all quantum optimization algorithms. The Qiskit Runtime enabled version of the Warm-start QAOA performed identically to the normal implementation

Table 4: Unsupervised Clustering Results by Algorithm on IBM Lagos with Runtime Environment

| Simulator | Type | Classical | VQE (Runtime) | QAOA (Runtime) | ws-QAOA (Runtime) |
|---|---|---|---|---|---|
| Honda Civic | economy | 1 | 0 | 1 | 0 |
| Toyota Corolla | economy | 1 | 0 | 1 | 0 |
| Camaro Z28 | sport | 0 | 1 | 0 | 1 |
| Pontiac Firebird | sport | 0 | 1 | 0 | 1 |
| Maserati Bora | sport | 0 | 0 | 0 | 1 |
| Energy (Ha) | | -414.86 | -34.025 | -18.476 | -290.264 |
| Solution Objective | | 1163.497 | 0.0 | 0.0 | 1163.497 |
| Process time (s) | | 0.014 | 4.125 | 3.81 | 4.35 |

Table 5: Unsupervised Clustering Results by Algorithm on IBM Nairobi with the Qiskit Runtime Environment for wine data set

| Simulator | Class | Classical | VQE (Runtime) | QAOA (Runtime) | ws-QAOA (Runtime) |
|---|---|---|---|---|---|
| Wine 1 | 0 | 0 | 0 | 1 | 0 |
| Wine 2 | 1 | 1 | 0 | 1 | 1 |
| Wine 3 | 1 | 1 | 0 | 1 | 1 |
| Wine 4 | 0 | 0 | 0 | 1 | 0 |
| Wine 5 | 0 | 0 | 1 | 1 | 0 |
| Wine 6 | 1 | 1 | 0 | 1 | 1 |
| Energy (Ha) | | -6714.018 | -163.109 | -33.243 | -780.723 |
| Solution Objective | | 5394.010 | 1743.006 | 0.0 | 52229.009 |
| Process time (s) | | | 5.800 | 5.313 | 5.609 |

of Warm-start QAOA.

On the IBM Lagos seven qubit quantum computer, the results show that Warm-start QAOA performed significantly better than normal VQE or QAOA. Warm-start QAOA correctly clustered all observations in the data set when run on quantum hardware, and provided solution objective scores that were equivalent to the classical algorithm. It is interesting to note that warm-start QAOA performed equivalently in terms of assigning observations to the correct cluster as well as in terms of solution objective for the Qiskit Runtime and normal implementation of Warm-start QAOA. In addition, Qiskit Runtime enabled warm-start QAOA ran 18 times faster than the normal implementation of warm-start QAOA, providing similar results in a fraction of the time. It should be noted that warm-start QAOA took nearly three times as long as normal QAOA when not run using Qiskit Runtime, making this methodology significantly more costly in terms of computer time when not run using Qiskit Runtime.

## 4 Discussion and Conclusions

Our results demonstrate that warm-start QAOA produces more consistent results than other algorithms on quantum simulators. This work also showed that non-convex quadratic programs can be relaxed successfully using the GuRoBi optimizer and used to make QAOA consistently produce consistently better performance.

In addition, our study also showed that Qiskit Runtime does not negatively impact the performance of quantum optimization algorithms on a quantum simulator, and in the case of warm-start QAOA produced equivalent results 18 times faster. The equivalent performance of warm-start QAOA running normally and through Qiskit Runtime shows Qiskit Runtime can provide equivalent performance in a fraction of the time when quantum algorithms use a warm-start procedure. Neither VQE or normal QAOA provided correct clustering results when run on either a simulator, quantum hardware, or with Qiskit Runtime enabled. Warm-start QAOA produced correct clustering results and solution objective score equivalent to the classical algorithm regardless of whether run on the simulator, on quantum hardware, or with Qiskit Runtime enabled. Warm-start QAOA consistently showed the best performance of the quantum optimization algorithms for clustering tasks.

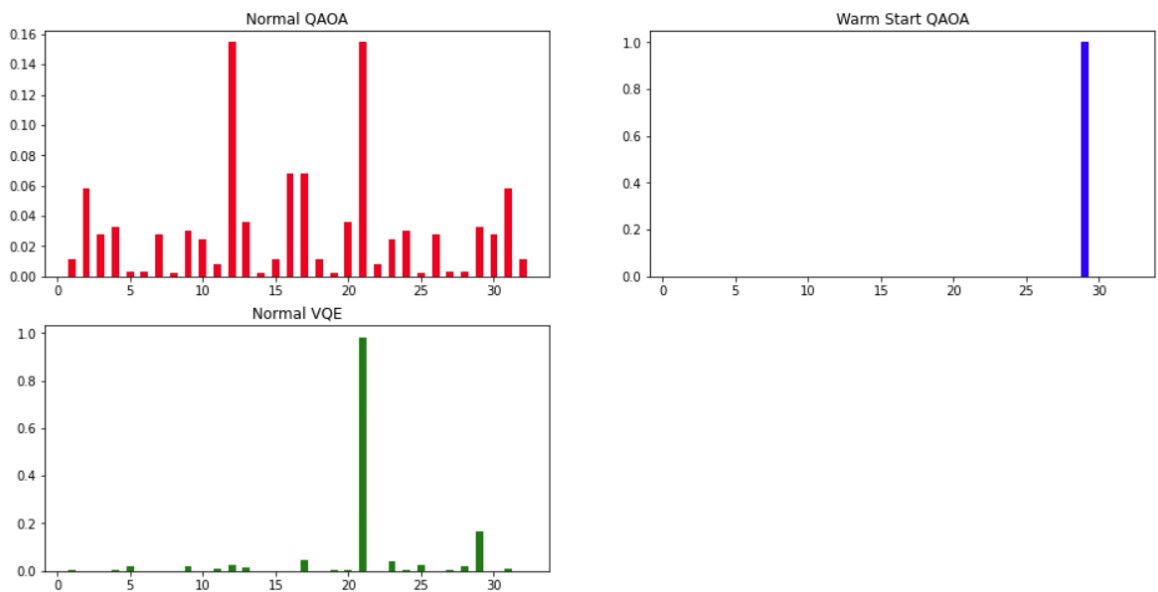

Figure 3: Solutions for Quantum Optimization Algorithms on Simulator

Overall the results indicate that quantum algorithms can consistently produce unsupervised clustering results when using the MAXCUT formation that are as good as a classical machine when using the warm-start QAOA methodology on current low qubit NISQ systems. Qiskit Runtime is a positive step towards speeding up quantum computation. The finding that Qiskit Runtime sped up warm-start QAOA is especially important since warm-start QAOA took three times as long as normal QAOA to run on quantum hardware. Graph based unsupervised clustering is hypothesized to require fewer iterations to reach a optimal answer than classical computational algorithms [1] [2]. When the quantum computing performance and quantum volume increase, it is theoretically possible that quantum computers may be able to perform unsupervised clustering faster than classical computers.

## 5 Acknowledgement

We would like to thank Jessie Yu, Paul Kassebaum and Julien Gacon for helpful comments and discussion.